\documentclass[aps,prl,twocolumn,showpacs,superscriptaddress]{revtex4}
\usepackage{epsfig}
\usepackage{color}
\usepackage{ulem}
\newcommand{\tr}[1]{\textcolor{black}{#1}}
\newcommand{\tb}[1]{\textcolor{black}{#1}}
\newcommand{\tg}[1]{\textcolor{black}{#1}}
\newcommand{\ave}[1]{\ensuremath{\langle #1\rangle}}
\newcommand{\bra}[1]{\ensuremath{\langle #1 |}}
\newcommand{\ket}[1]{\ensuremath{|#1\rangle}}
\newcommand{\be}{\begin{equation}}
\newcommand{\ee}{\end{equation}}
\newcommand{\ba}{\begin{eqnarray}}
\newcommand{\ea}{\end{eqnarray}}
\def\lsim{\mathrel{\rlap{\lower4pt\hbox{\hskip0pt$\sim$}}
    \raise1pt\hbox{$<$}}}
\def\gsim{\mathrel{\rlap{\lower4pt\hbox{\hskip0pt$\sim$}}
    \raise1pt\hbox{$>$}}}
\DeclareMathSizes{10}{10}{5}{3}

\begin{document}

\title{\tb{Photons walking the line: A quantum walk with adjustable coin operations}}

\author{A. Schreiber}  
\author{K.N. Cassemiro}
   \email{Katiuscia.Cassemiro@mpl.mpg.de}
\affiliation{Max Planck Institute for the Science of Light, G\"{u}nther-Scharowsky-str. 1 / Bau 24, 91058 Erlangen, Germany.}
\author{V. Poto\v cek}
\affiliation{Department of Physics, FNSPE, Czech Technical University in Prague, B\v rehov\'a 7, 115 19 Praha, Czech Republic.}
\author{A. G\'abris}
	\altaffiliation{\tb{Secondary address: Research Institute for Solid State Physics and Optics, Hungarian  Academy of Sciences, H-1525 Budapest, P. O. Box 49, Hungary.}}
\affiliation{Department of Physics, FNSPE, Czech Technical University in Prague, B\v rehov\'a 7, 115 19 Praha, Czech Republic.}    
\author{P.J.Mosley}
   \altaffiliation{Present address: Centre for Photonics and Photonic Materials, Departmen of Physics,  University of Bath, Claverton Down, Bath, BA2 7AY, United Kingdom.}
\affiliation{Max Planck Institute for the Science of Light, G\"{u}nther-Scharowsky-str. 1 / Bau 24, 91058 Erlangen, Germany.}
\author{E. Andersson}
\affiliation{SUPA, School of EPS, Heriot-Watt University, Edinburgh EH14 4AS, United Kingdom.}
\author{I.Jex}   
\affiliation{Department of Physics, FNSPE, Czech Technical University in Prague, B\v rehov\'a 7, 115 19 Praha, Czech Republic.}
\author{Ch. Silberhorn}
\affiliation{Max Planck Institute for the Science of Light, G\"{u}nther-Scharowsky-str. 1 / Bau 24, 91058 Erlangen, Germany.}

\date{\today}

\begin{abstract}
We present the first robust implementation of a coined quantum walk over five steps using only passive optical elements. By employing a fiber network loop we keep the amount of required resources constant as the walker's position Hilbert space is increased.  We observed a non-Gaussian distribution of the walker's final position, thus characterizing a faster spread of the photon wave-packet in comparison to the classical random walk.  The walk is realized for many different coin settings and initial states, \tb{opening} the way for  the implementation of  a quantum walk-based search algorithm.
\end{abstract}


\maketitle

Random walks are one of the fundamental models of natural sciences. The concept is common to many branches of research,  for example describing material transport in media and the evolution of stock market shares ~\cite{1}. By endowing the walker with quantum properties many new interesting effects appear. As first noted by Aharonov {\it et al.}~\cite{5}, quantum interference leads to a new type of walk that spreads much faster than its corresponding classical counterpart. Since classical walks constitute a computational primitive, it can be expected that their quantum extensions provide an alternative platform for the implementation of quantum information protocols. It has  been theoretically proven that quantum walks allow the speed-up of search algorithms~\cite{qrw_search,search_vasek} and the realization of universal quantum computation~\cite{qrw_computation}. Moreover, they can be employed for testing the transition from the quantum  to the classical world by applying a controlled degree of decoherence~\cite{30}. In the context of time-dependent phenomena, recent theoretical studies of quantum walks with a sufficiently large number of sites have shown highly non-trivial dynamics, including localization and recurrence~\cite{recurrence}. Applying such ideas, for example to a biophysical system, can give important insights into effects like photosynthesis~\cite{11}.

While theoretical analysis of quantum walks is advanced, only few experiments have been reported. The system chosen for implementation has to allow for quantum interference and maintain coherence for a sufficiently long time. To date, different experimental approaches have been taken. Several steps of a quantum walk were realized with trapped ions or atoms~\cite{12}.  Taking advantage of the simple preparation and manipulation of light states~\cite{17}, the recent realization of quantum walks with photons has also attracted attention~\cite{13}. In this Letter we report on the implementation of a one-dimensional coined quantum walk based on optical networks, which corresponds to a quantum analogue of a Galton board. While our primary aim is a demonstration of the ex\-pe\-rimental feasibility with a low degree of decoherence, the employed configuration is scalable in terms of reachable number of steps and accessible position Hilbert space. In contrast to previous implementations, we designed a setup for an optical implementation of the coined quantum walk, which presents the distinct advantage of high flexibility in the manipulation of the walker's internal degree of freedom.

In our implementation we exploit the polarization of the photon as internal degree of freedom, which can be described with the basis states $\ket{H}=(1,0)^{T}$ and $\ket{V}=(0,1)^{T}$. In the elementary version, the quantum walker performs a spatial shift (step) conditioned on its internal state.  If the motion is restricted to a line, the shift occurs either to the left or to the right and the resulting position is represented by integer values $x$. In mathematical terms, one step of the quantum walk is determined by the product of two unitary operators. After $n$ steps, the evolution operator $\hat{U}$ is given by $\hat{U}=(\hat{S}\,\hat{C})^{n}$, with
\begin{equation}
\hat S = \sum_x \vert x-1\rangle \langle x\vert\otimes\ket{V}\bra{V}+ \vert x+1\rangle \langle
x\vert\otimes \ket{H}\bra{H},
\end{equation}
describing the spatial shift  (step operator), and $\hat{C}$ the tossing of the quantum coin, which operates on the polarization of the photon (see below). The coherent action of the step and coin operators leads to entanglement between the position and the internal degree of freedom. After several steps, the counterintuitive profile of the quantum walk probability distribution  emerges as a result of quantum interference among multiple paths.

Despite the appeal of performing a quantum walk using only linear optical components, a straightforward implementation of the Galton board requires the use of multiple beam splitters and phase-shifters~\cite{22}. This increases the experimental complexity in terms of optical stability, alignment and cost. In our implementation we circumvent this obstacle by translating the position of the walker (photon) into arrival times at the detector. Since the coin operator acts on the polarization subspace, it is simply implemented using a half-wave plate (HWP). Its matrix representation on the \{\ket{H}, \ket{V}\} basis is
\begin{equation}
 C =    \left(
                 \begin{array}{rr}
                           \cos(2 \theta) & \sin(2 \theta)  \\
                           \sin(2 \theta) & -\cos(2 \theta) \\
                 \end{array}
               \right) ,
\end{equation}
\tr{where $\theta$ is} the rotation angle of the HWP relative to one of its optical axes.
The evolution of the walk is perpetuated using an optical feedback loop~\cite{23}, which allows us to completely avoid the use of additional optical elements to realize several steps of the walk. Similar ideas employing optical networks have been applied with considerable success in other experiments for obtaining a time-multiplexed detector~\cite{25}. Here we advance this concept significantly by realizing a network that includes interferences among multiple paths.

\begin{figure}[ht]
\centering \epsfig{file=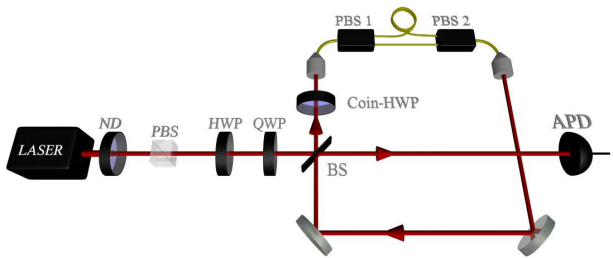,scale=0.21}
\caption{(color online). Sketch of the setup. A laser field is attenuated to the single-photon level via neutral density filters (ND) and coupled into the network loop through a 50/50 beam splitter (BS). HWP: half-wave plate; QWP: quarter-wave plate;  PBS: polarizing BS; APD: avalanche photodiode. Setup dimensions: 1.5~m in free space and 7~m (8~m) in fiber when horizontal (vertical) polarization is used.} 
\label{setup}
\end{figure}

\begin{figure}[ht]
\centering 
\epsfig{file=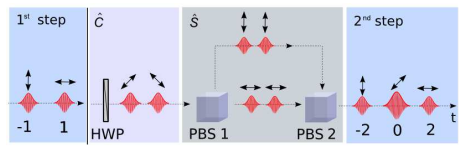,scale=0.26}
\caption{(color online). Illustration of the working principle of the setup. From left to right the walk evolves from the end of the first step to the end of the second. The arrows represent the polarization of the photon, which was assumed to be  initially vertical (``zero'' step).} 
\label{principle}
\end{figure}

Our experimental setup is sketched in Fig.~\ref{setup}. The photon's wave packet is provided by a pulsed laser source with central wavelength of 805~nm, pulse width of 88~ps and repetition rate of 1~MHz. The pulses are attenuated to the single photon level by using neutral density filters. The initial polarization is prepared using standard half- and quarter-wave plates, the coin is realized by another HWP, and the step operation by an optical feedback loop. The ``stepper'' is composed of a polarization-maintaining fiber network, such that the horizontal and vertical components are first separated spatially and then temporally in a deterministic way. 
Horizontally polarized photons traverse the fiber loop network in 40~ns, while vertical ones take 5~ns longer. At the output of the ``stepper'' the two paths are coherently recombined and the photon is sent back to the input beam splitter for the next step. An illustration of the evolution of the photon's wave packet through the optical network is shown in Fig.~\ref{principle}.
At first inspection it seems that no interference can occur due to the orthogonal nature of the states that are recombined at the end of the fiber. Nevertheless, in the next iteration, the coin operation creates a superposition of the states, thus displaying interference when analyzing in the $\ket{H}$ and $\ket{V}$ basis (PBS 1 on Fig.~\ref{setup}). Finally, at each step of the walk (corresponding to one loop) there is a 50\% probability of coupling the photon out of the loop, in which case an avalanche photodiode (APD with time jitter $<$1~ns) will register a click. The detection efficiency is $\eta_{\mathrm{det}}=0.24(1)$ and the losses in the setup are characterized by an efficiency of $\eta_{\mathrm{setup}}=0.18(1)$ per step, neglecting the input coupling. Measurements of the transmitted (and/or reflected) count rates after each optical component enable us to characterize the losses for $\ket{H}$  and $\ket{V}$ polarizations. A glass plate was introduced in the setup to equalize the losses; nevertheless  $\ket{H}$ polarized photons experienced 3\% less losses per step than $\ket{V}$ ones. 

The characterization of the walk consisted of a series of consecutive runs of the experiment, each generating at most a single click at a specific time, which is recorded by a computer via a time-to-digital converter interface. From the technical point of view, we stress that 
phase stability is required only during the short time scale of a single experiment (e.g. 225~ns for five steps), in contrast to the longer time required for an ensemble measurement. This fact brings the advantage that no active phase stabilization was required in our experiment.

To demonstrate the crucial properties of our implementation we conducted two different types of measurements. In the first experiments we show a high degree of coherence and the scalability of the system by studying the evolution of the walk over five steps. The second set of experiments show the flexibility achieved for the manipulation of the coin. We begin with the study of coherence properties over an increasing number of steps. The probability distribution of the quantum  walk is highly sensitive to the initial state. The best way to emphasize the differences between the quantum and the classical walk\textendash or in other words to test for coherence\textendash is to use a particular balanced input state, i.e. a circularly polarized photon. In addition, by using the Hadamard coin ($\theta=22.5^{\circ}$), which creates an equal superposition of horizontal $\ket{H}$  and vertical $\ket{V}$ polarizations, the wave packet of the photon evolves into a highly delocalized state.

\begin{figure}[ht]
\centering 
\includegraphics[scale=0.38]{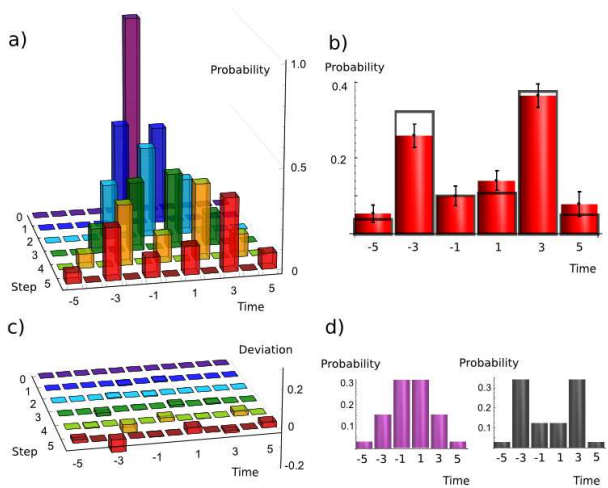} 
\caption{(color online).
Measured probability distribution of the photon's arrival time. (a) Evolution of the distribution from the initial circularly polarized state (rear part) to the state after the fifth step (front part). (b)  Detail of the measured distribution after five steps.  Filled bars: measured results. Frames: predictions from our theoretical model.  (c) Difference between experimental and theoretical values.  (d)  Detail of ideal distributions after five steps. Left: classical walk; right: quantum walk.} 
\label{fivesteps}
\end{figure}

We prepared the initial circular polarization state $\ket{p}_i = a_H\ket{H}+ e^{i \Phi} a_V\ket{V}$ with an accuracy characterized by the factor $|a_H|^2/|a_V|^2=0.94(4)$, yielding a fidelity of $F=99.9\%$. The initial mean photon number was $\ave{n}_{\mathrm{initial}}=8(2)$  and, after the fifth step, $\ave{n}_{\mathrm{5step}}\approx 7\cdot10^{-4}$. The measured evolution of the probability distribution for the photon's arrival time from the first to the fifth step is shown in Fig.~\ref{fivesteps}a. 
\tg{Here one can see the gradual decrease of the probabilites of a photon arriving in the central time bins alongside the growth of arrivals in the outer wings~\textendash~a distinctive feature of the quantum walk}. This is a clear signature of a high degree of coherence throughout the complete evolution of the walk.

The delocalization effect can be better appreciated in Fig.~\ref{fivesteps}b, in which we show the measured distribution after five steps. In addition to our experimental data we present a comparison with both  the theoretical model applied to our setup (Fig.~\ref{fivesteps}b-c) and the classical Gaussian distribution (Fig.~\ref{fivesteps}d). While in the classical case the standard deviation is given by $\sigma_C=\sqrt{5}\approx 2.24$, a higher deviation occurs in the quantum case ($\sigma_{\mathrm{Q}}=2.83$). Our measured value $\sigma=3.03(10)$ agrees well with the expected ballistic spread. The presented error bars include only  statistical errors, calculated as the standard deviation of the finite number of experiments (N=3016 in the fifth step). By analyzing the evolution of the walk over several steps we find that the current limitation in implementing more than 5 steps arises from spurious optical reflections. Those can be largely supressed by appropriate time gating, but they still lead to a systematic error in the probability distribution. In addition, the use of the 50/50 beam-splitter (BS) coupler introduces high losses, which in turn causes a low signal to noise ratio at the detection of further steps. We stress that these problems are not intrinsic to this implementation, since the setup can be optimized to give a better performance (discussed below).

\begin{figure}[ht]
\centering \epsfig{file=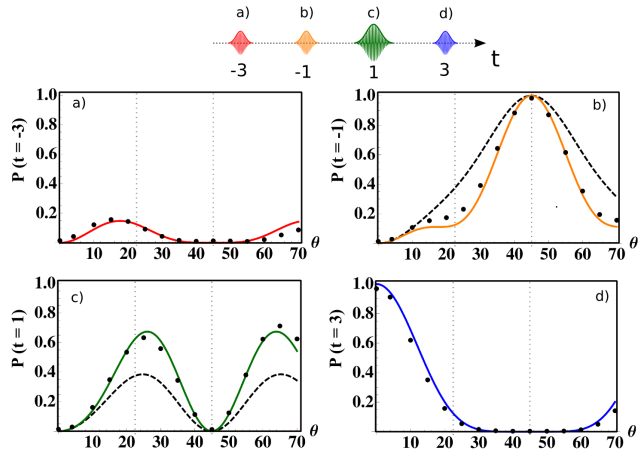,scale=0.82} 
\caption{(color online).
Effect of different coin operations on the probability distribution after the third step for an initially horizontally polarized photon. In each inset, from (a) to (d), we show the probability for the photon to arrive at a particular time bin, as depicted in the top illustration. At time bins $t = \pm3$ the classical and quantum descriptions coincide. Dots:  Measured quantum  walk (error bars are smaller than used symbols).  Solid line: Theoretical model for the quantum walk. Dashed line: Classical random walk.} 
\label{changecoin}
\end{figure}

Our second experimental result highlights the flexibility of our implementation with respect to the easy adjustability of different coin settings. In Fig.~\ref{changecoin} we show how the probability distribution after three steps changes as a function of the angle of the half-wave plate.  In this case the photon is initially prepared with horizontal polarization, leading to an asymmetric distribution when the Hadamard coin is applied. Setting the HWP at zero degrees is essentially equivalent to applying the identity operation, thus resulting in the photon being found at the first time bin, labelled  $t=3$. From zero to $45^{\circ}$ interference among multiple paths takes place, giving rise to a probability of finding the photon at $t=1$ ($t=-1$) that is increased (decreased) in comparison to the classical result. At exactly $45^{\circ}$ the NOT operation is realized, i.e. 
incoming $\ket{H}$ polarized photons are converted into  $\ket{V}$ and vice-versa. For these measurements, the initial polarization state $\ket{p}_i = a_H'\ket{H}+ a_V'\ket{V}$ was characterized by  $|a_V'|^2/|a_H'|^2=0.003(4)$ ($F=99.7$) and the initial mean photon number was given by $\ave{n}=0.58(5)$. The experimental results agree well with the theoretical prediction and are clearly distinct from the classical values. 

We performed a detailed theoretical analysis of the system by taking into account possible sources of coherent and incoherent errors. The imperfections are modeled as additional linear optical elements and represent the effect of depolarization, relative phase shifts and efficiency ratio $\epsilon$ between the two polarizations, undesired polarization rotations by an angle $\varphi$, and imperfect preparation of the initial state $\ket{\psi '}$. By analyzing the output signal, the strength of decoherence has been found to be equal to zero within the statistical error, indicating that effects such as depolarization and rapid phase fluctuations can be neglected. The system can be described by an effective coin operation $\hat{C}'$, with \tr{matrix} representation given by
\begin{equation}
C'= L(\epsilon_{L}) R(\varphi) R(\theta) L(-\epsilon_{\mathrm{HWP}}) R(-\theta) L(\epsilon_{\mathrm{BS}})\, ,
\end{equation}
where $L(\epsilon)$ is a matrix characterizing differential losses and $R(\alpha)$ a rotation:
\be
L(\epsilon ) =  \left(
                 \begin{array}{rr}
                           1 & 0  \\
                           0 & \epsilon \\
                 \end{array}
               \right),~~R (\alpha)=
  \left(
                 \begin{array}{rr}
                           \cos(\alpha) & \sin(\alpha) \\
                         - \sin(\alpha) & \cos(\alpha) \\
                 \end{array}
               \right) .
\ee
Values $\epsilon < 1$ for the efficiency ratio indicate loss \tr{imbalance} between the $\ket{H}$ and $\ket{V}$ polarizations. This parameter was characterized at the different components of the setup: at the coupling beam splitter $\epsilon_{\mathrm{BS}}=0.99$, at the delay loop $\epsilon_L =0.96$, and between the slow and the fast axis of the HWP $\epsilon_{\mathrm{HWP}}=0.98$ . The results indicate that $\ket{V}$ polarization undergoes higher losses than $\ket{H}$. The rotation introduced by the mirrors has been determined to be $\varphi =1.4^{\circ}$. The final state of the walk is calculated as $\ket{\psi_{\mathrm{final}}}=(\hat{S}\hat{C}')^n \ket{\psi '}$.

Discrepancies between the experiment and the theoretical model are due to reflections of the optical signal and imperfections of the detector, e.g. dead time and dark counts. Considering only intrinsic problems of the actual proposal, i.e. supposing the use of the best available components, thereby suppressing the reflections, we can estimate the maximum number of steps that are in principle possible to achieve. In the optimized scenario it is reasonable to suppose $\eta_{\mathrm{setup}}=$71\%, considering that the 50/50 coupler is replaced by a 99/1 coupler. Since the properties of the quantum walk can be simulated by using an intense coherent field, we could employ a laser with 1~W power (250~kHz). By adding an active switch to couple the photon out of the loop, the signal to noise ratio is improved, thus allowing to reach 100 steps. The measurements can still be done in a time scale shorter than typical unavoidable low frequency mechanical vibrations (lower than 500Hz), thus preserving phase coherences.

In conclusion, we have implemented a compact and efficient way of realizing coined quantum walks. In contrast to many other experiments, we benefit from employing quantum states of light, which are simple to manipulate. For instance, one could achieve a higher dimensional coin by using the optical angular momentum of photons~\cite{27} instead of their polarization, therefore increasing the dynamical richness of the walk. Moreover, the ability to operate with different coins and the ease of addressing individual position states opens exciting new possibilities for the realization of quantum information protocols. The present experimental setup constitutes a starting point for implementing a one- or  two-dimensional quantum walk-based search algorithm.


We acknowledge financial support from the German Israel Foundation (Project 970/2007). 
K.N.C. and I.J. acknowledge financial support from the Alexander von Humboldt Foundation; V.P., A.G. and I.J.
from M\v SMT LC06002, MSM 6840770039 and CZ-10/2007; V. P. from GA CR 202/08/H078; A.G. from the Hungarian Scientific
Research Fund (T049234 and NF068736); E.A., I.J., V.P. and A.G.  from the Royal Society International Joint Project grant 2006/R2--IJP. 

\end{document}